\pdfoutput=1
\documentclass[a4paper,11pt]{article}
\usepackage{pos}
\usepackage{hyperref}

\title{CTbend: A Bayesian open-source framework to model pointing corrections for Cherenkov telescopes}
 \ShortTitle{CTbend: Pointing corrections for Cherenkov telescopes}

\author*[a]{Gerrit Spengler}
\author[a]{Ullrich Schwanke}
\author[b,c]{Dmitriy Zhurov}

\affiliation[a]{Institut f\"ur Physik, Humboldt-Universit\"at zu Berlin, Newtonstr.
15, 12489 Berlin, Germany}

\affiliation[b]{Applied Physics Institute of Irkutsk State University (API
ISU), Irkutsk, Gagarina Boulevard 20, Russian Federation}
\affiliation[c]{Irkutsk National Research Technical University (INRTU),
Irkutsk, St. Lermontov 83, Russian Federation}

\emailAdd{spengler@physik.hu-berlin.de}
\emailAdd{schwanke@physik.hu-berlin.de}
\emailAdd{sidney28@yandex.ru}

\abstract{The pointing of Cherenkov telescopes is subject to imperfections which are, for example, related to
the bending of the telescopes mechanical structure. These imperfections must be measured, modeled, and finally corrected to
achieve an optimal telescope pointing precision. The measurement of pointing deviations is often 
performed while the telescope points to different stars and a CCD camera
monitors the offsets of the star images to the center of the focal plane. Outliers in these
measurements can propagate into the pointing model and lead to imprecise model predictions.
CTbend is a simple and standalone open-source framework that uses a Bayesian
analysis with an outlier resilient likelihood function to model the pointing of Cherenkov
telescopes with parametric standard models like TPoint. The framework is described in the following.}
\FullConference{37$^{\rm{th}}$ International Cosmic Ray Conference (ICRC 2021)\\
		July 12th -- 23rd, 2021\\
		Online -- Berlin, Germany}

\begin{document}
\maketitle

\section{Introduction}
\label{introduction}
Cherenkov telescopes are routinely used to detect very high-energy gamma-rays. For example, the HESS array of Cherenkov telescopes, which is operated in Namibia,
is sensitive to gamma-rays with energies in the range of a few 10 GeV up to a few 100 TeV \cite{hess}.
Other currently operating examples for arrays of Cherenkov telescopes are TAIGA, MAGIC and VERITAS \cite{taiga,magic,veritas}.
It is planned that the Cherenkov Telescope Array (CTA) will start operations within few years \cite{cta}.\\
Mechanical imprecisions in the construction and gravitational effects lead to telescope mispointing.
This mispointing depends on the azimuth and elevation angle of the horizontal telescope pointing.
CTbend is an open source\footnote{Source code and an example application based on simulated tracking data are available at \url{https://github.com/residualsilence/ctbend}.}
python package to model pointing corrections for Cherenkov telescopes which is described in the following. Input "pointing run data", 
used to derive pointing models in CTbend, is detailed in Sec. \ref{pointing_run_data}. Geometric nomenclatures are introduced
in Sec. \ref{geometry_section} and the derivation of pointing models is discussed in Sec. \ref{pointing_model_section}. Finally,
an outlook on future work is discussed in Sec. \ref{outlook_section}.

\section{Acquisition of pointing run data}
\label{pointing_run_data}
\begin{figure}[!tbp]
	\begin{minipage}[b]{0.48\textwidth}
	\includegraphics[width=\textwidth]{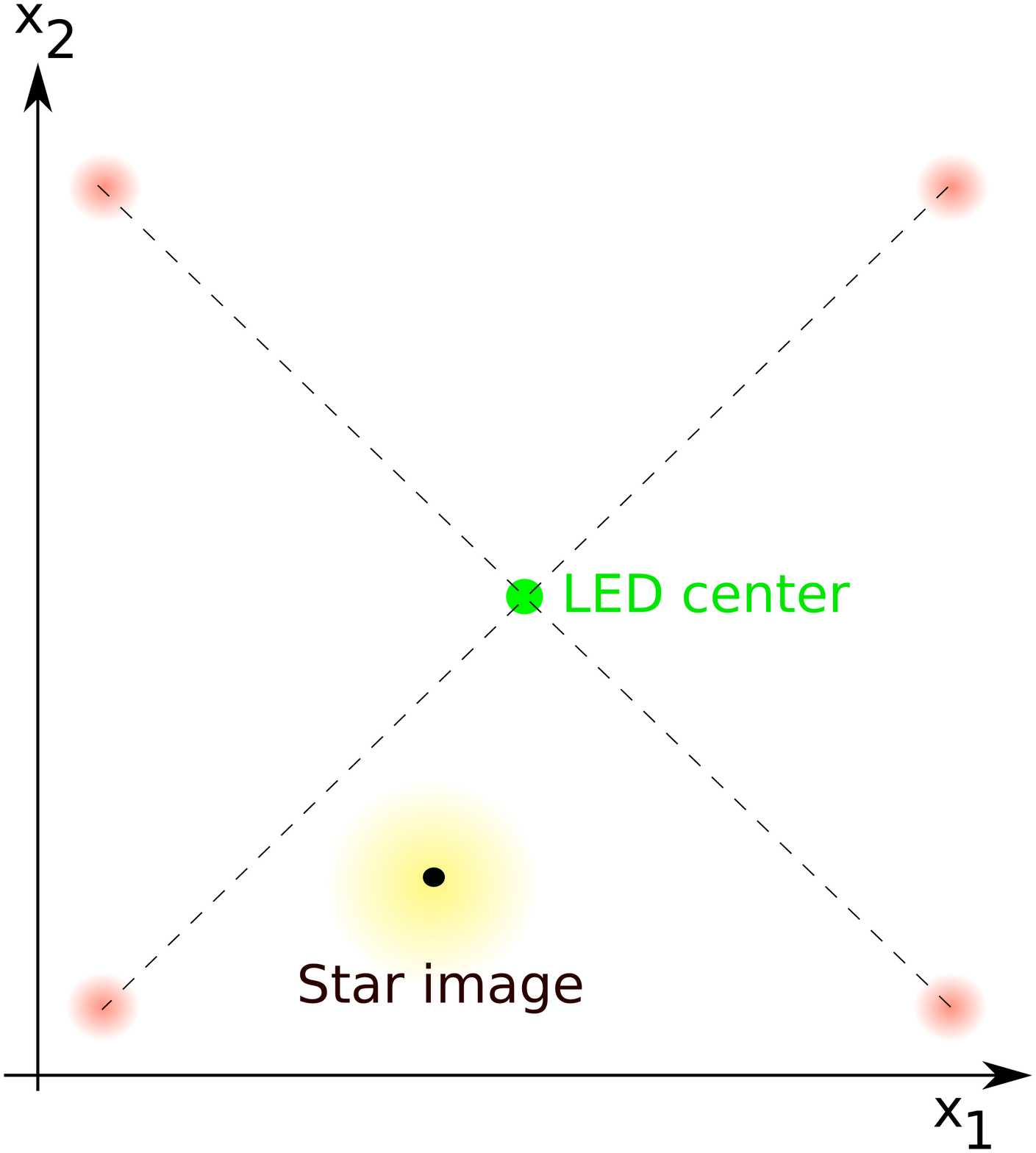}
	\caption{Sketch of the focal plane as observed by a CCD camera installed in the
center of the telescope dish. The $x_1-x_2$-coordinate
system is given by the pixel matrix of the CCD image. Shown in red are the
images of positional LEDs whose center (in green) defines the pointing direction
of the telescope. Shown in yellow is the image of a star to which the telescope
is pointed. The pointing model predicts the displacement of the center of the
star image from the telescope pointing direction as a function of the telescope
azimuth and elevation.}
\label{focalplane_sketch}
	\end{minipage}
	\hfill
	\begin{minipage}[b]{0.48\textwidth}
		\includegraphics[width=\textwidth,angle=-90]{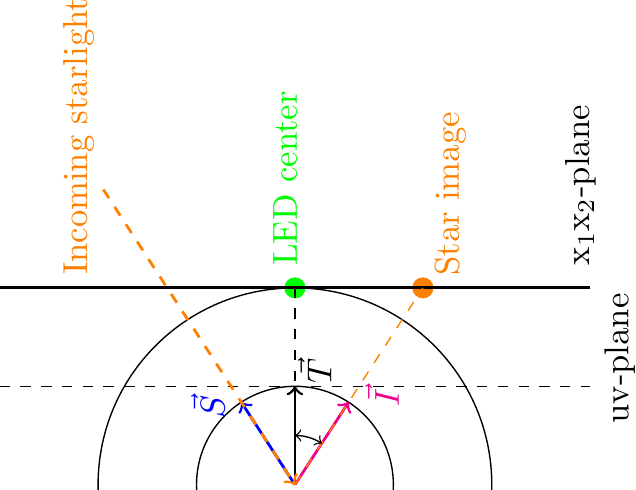}
	\caption{Geometric connection between $\mathrm{x_1}\mathrm{x_2}$- and uv-coordinates. $\vec{T}$ is a unit vector towards the horizontal telescope coordinates. 
	$\mathrm{x_1}\mathrm{x_2}$-coordinates describe the focal plane as imaged by a CCD camera. 
	The uv-plane is spanned by $(\vec{e_{\phi D}},\vec{e_{\theta D}})$, i.e. it is the tangential plane on the unit vector $\vec{T}$. 
	The telescope is supposed to point towards a star, i.e. towards the unit vector $\vec{S}$. 
	The actual telescope direction is inferred from the center of an LED pattern on the focal plane, indicated by a green point. 
	The vector $\vec{I}$ is the vector $\vec{S}$ reflected on the telescope direction $\vec{T}$ 
	and points towards the image of the star in the $\mathrm{x_1}\mathrm{x_2}$-plane. It holds $\protect\angle (\vec{T},\vec{I})=\protect\angle (\vec{S},\vec{T}):=\psi$.
	}
	\label{geometry_fig}
	\end{minipage}

\end{figure}
An example for the acquisition of pointing run data, i.~e. data from which a pointing model for a Cherenkov telescope can be derived, is described in this section. 
It is assumed that the focal plane of the telescope is observed with a CCD camera which is mounted at the telescope dish. 
Figure \ref{focalplane_sketch} shows a simplified sketch of a CCD-image of the focal plane in which positional LEDs are arranged as square, 
the centre of which is defined as the telescope pointing direction. Consider that the telescope is pointed towards a bright star. 
Without bending effects, the center of the star image in the focal plane and the telescope pointing direction are expected to be spatially coincident. 
However, bending effects lead to a mispointing, which in turn leads to a displacement between the pointing direction and the center of the star image.\\
In the following, $(x_1,\,x_2)$ denote Cartesian coordinates which describe positions of pixels in a CCD image. 
The mispointing between the actual and the requested telescope direction is measured as displacement between the 
telescope pointing direction 
$(x_1^\mathrm{tel},\, x_2^\mathrm{tel})$ and the direction of a star $(x_1^\mathrm{star},\,x_2^\mathrm{star})$. Figure \ref{focalplane_sketch} illustrates this nomenclature.\\
Pointing run data is a collection of $2N$ displacements $\Delta x_{1i}:=x_{1i}^\mathrm{star}-x_{1i}^\mathrm{tel}$ and $\Delta x_{2i}:=x_{2i}^\mathrm{star}-x_{2i}^\mathrm{tel}$ 
where $i=1\dots N$ correspond to data taken at a representative sample of pointing directions in the horizontal coordinate system. 
A pointing model predicts the telescope mispointing in horizontal coordinates and in turn the 
displacements $\Delta x_1,\,\Delta x_2$ as a function of the requested telescope direction to enable the correction of the mispointing.\\ 
It must be expected in practice that outlier occur in pointing run data, i.~e. large displacements $(\Delta x_1,\,\Delta x_2)$ 
which are unrelated to bending effects. These outliers can have multiple origins, examples are an inhomogeneous night sky background, CCD pixel errors or undesired reflections.

\section{Geometric description of mispointing}
\label{geometry_section}
\subsection{Horizontal coordinates}
Consider a horizontal coordinate system where the azimuth angle is math negative, i.~e. the azimuth angle $\phi$ 
is getting smaller when the telescope is moving on the shortest path from south to east\footnote{For example: South is at $\phi=0$ and east is at $\phi=-90^\circ$.}, 
and an elevation of $\theta=0$ corresponds to the direction of the horizon. Unit vectors in this coordinate system are given by
\begin{equation}
	\vec{V}(\phi,\theta)=\left(\begin{array}{c}\cos\theta\cos\phi\\ -\cos\theta\sin\phi \\ \sin\theta\end{array}\right)\;\mathrm{.}
\end{equation}
In the following, it is assumed that $\vec S=\vec V(\phi,\,\theta)$ is the direction of a bright star. 
To observe towards $\vec S$, the telescope 
drive system is commanded towards $\phi_D:=\phi+\Delta\phi_0$ and $\theta_D:=\theta+\Delta\theta_0$. The correction $(\Delta\phi_0,\,\Delta\theta_0)$ is the applied pointing model and 
aims to compensate for the telescope mispointing. 
When the telescope drive system is commanded towards $(\phi_D,\,\theta_D)$, the actual telescope pointing direction is given by $\vec T=\vec V(\phi_D-\Delta\phi,\,\theta_D-\Delta\theta)$. An optimal 
pointing is achieved when the applied pointing correction $(\Delta\phi_0,\,\Delta\theta_0)$ equals the actual mispointing $(\Delta\phi,\,\Delta\theta)$.\\
In general, the starlight is reflected at the telescope dish and imaged towards
\begin{equation}
	\label{star_image}
	\vec{I}:=2(\vec{S}\cdot\vec{T})\vec{T}-\vec{S}\,\mathrm{,}
\end{equation}
as illustrated in Fig. \ref{geometry_fig}.
\subsection{The $uv$-coordinate system}
Consider the projected quantities
\begin{equation}
	\label{measurement_uv}
\begin{array}{l}
	u = \tan(\xi\mathrm{\tilde{\Delta x}_1})\\
	v = \tan(\xi\mathrm{\tilde{\Delta x}_2})
\end{array}
\end{equation}
where $\xi$ is the CCD camera pixel-scale and $(\mathrm{\tilde{\Delta x}_1},\mathrm{\tilde{\Delta x}_2})$ coordinates are the CCD coordinates of the 
star image rotated by an angle $\alpha$ around the CCD coordinates ($\mathrm{x_1^{tel}}, \mathrm{x_2^{tel}}$) of the telescope pointing direction,
\begin{equation}
	\label{rotation}
	\begin{pmatrix} \mathrm{\tilde{\Delta x}_1} \\ \mathrm{\tilde{\Delta x}_2} \end{pmatrix} = 
	\begin{pmatrix} \cos\alpha & \sin\alpha \\ -\sin\alpha & \cos\alpha \end{pmatrix} 
			\begin{pmatrix} \Delta x_1 \\ \Delta x_2 \end{pmatrix}\;\mathrm{.}
	\end{equation}
The angle $\alpha$ describes a possible rotation of the CCD pointing axis around the telescope pointing direction. 
The quantities $u$ and $v$ are distances on the tangential plane at the telescope pointing unit vector, see also Fig. \ref{geometry_fig}. 
The uv-plane is the tangential plane on $\vec{T}$ 
spanned by the unit vectors in the directions of the change of the telescope pointing direction with the telescope drive system 
parameters $(\phi_D,\theta_D)$,
\begin{equation}
	\vec{e_{\phi D}}=\left\|\frac{\partial\vec{T}}{\partial\phi_D}\right\|^{-1}\frac{\partial\vec{T}}{\partial\phi_D}
	\label{e_phi}
\end{equation}
\begin{equation}
	\vec{e_{\theta D}}=\left\|\frac{\partial\vec{T}}{\partial\theta_D}\right\|^{-1}\frac{\partial\vec{T}}{\partial\theta_D}\;\mathrm{.}
	\label{e_theta}
\end{equation}
The uv-plane is used later to perform the optimization of pointing model parameters.

\section{Pointing models in CTbend}
\label{pointing_model_section}
\begin{figure}
	\centering
	\includegraphics[width=0.7\textwidth]{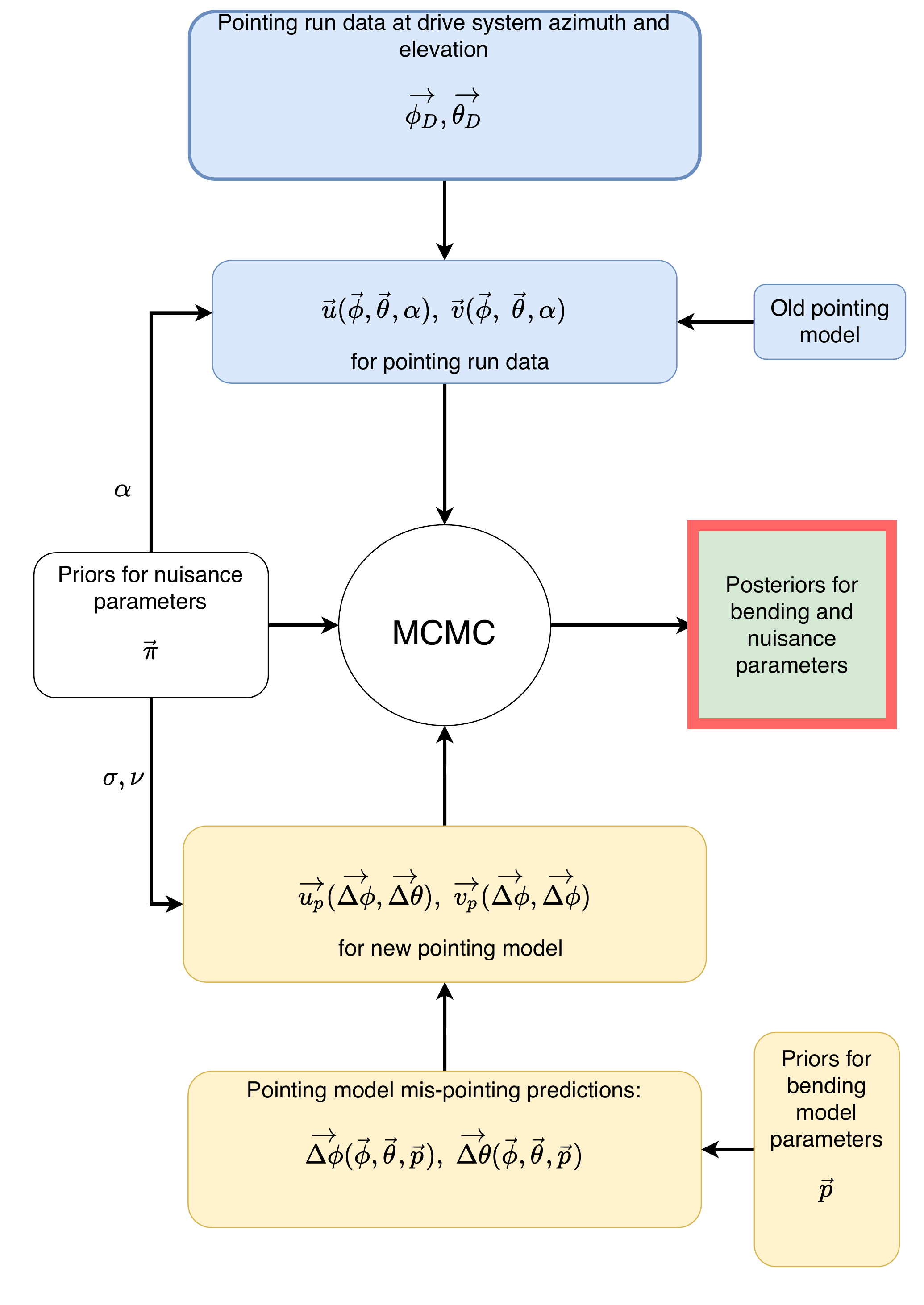}
	\caption{Analysis workflow: Given the likelihood function (Eq. \ref{likelihood_model}) and prior distributions for the pointing model and nuisance parameters, 
	the posterior parameter distributions are derived by matching the measured (blue boxes) and predicted (yellow boxes) values for $u$ and $v$ in a MCMC simulation.}
	\label{workflow_figure}
\end{figure}

Consider a parametric model 
\begin{equation}
	\begin{array}{l}
	\Delta\phi_p=\Delta\phi_p(\phi,\theta|\vec p,\vec\pi)\\
	\Delta\theta_p=\Delta\theta_p(\phi,\theta|\vec p,\vec\pi)
	\end{array}
\end{equation}
which predicts the telescope mispointing $\Delta\phi_p,\,\Delta\theta_p$ at horizontal coordinates 
$\phi,\,\theta$. The parameters $\vec p$ describe the pointing model while $\vec\pi$ are nuisance parameters, introduced by the measurement process for u and v. An example for a 
nuisance parameter is the CCD rotation parameter $\alpha$. Inspired by discussions in \cite{alma} and \cite{tpoint}, an example for the parameters of a linear model for the pointing correction 
is given in Tab. \ref{example_model_table} where
\begin{equation}
	\label{tpoint_model}
		\begin{array}{l}
			\Delta\phi_p=\sum_k\Delta\phi_{pk}\\
			\Delta\theta_p=\sum_k\Delta\theta_{pk}\\
			\mathrm{and}\\
			\vec p=(\mathrm{IA},\mathrm{IE},\mathrm{NPAE},\mathrm{AN},\mathrm{AW},\mathrm{TF},\mathrm{ACEC},\mathrm{ACES})\,\mathrm{.}

		\end{array}		
\end{equation}
Each parameter in $\vec p$ has a mechanical interpretation, many of which were already derived in \cite{meeks}. However, the model given by Eq. \ref{tpoint_model} is only 
an example and CTbend allows for a flexible implementation of arbitrary other parametric models.\\
The pointing model as well as nuisance parameters are to be optimized such that the predicted mispointings match with the actual telescope mispointings. 
Figure \ref{workflow_figure} summarizes the analysis workflow for the parameter optimization as implemented in CTbend and detailed in the following.

\subsection{Predicted mispointing in the uv-plane}
The prediction of a given pointing model is that the telescope is actually pointing to 
$\vec T_p=V(\phi_D-\Delta\phi_p\,\theta_D-\Delta\theta_p)$ when the drive system is commanded to $\phi_D=\phi+\Delta\phi_0$ and $\theta_D=\theta+\Delta\theta_0$ where $\Delta\phi_0$ and 
$\Delta\theta_0$ is the pointing model applied while taking data. Using Eq. \ref{star_image}, the vector $\vec I_p=2(\vec S\cdot\vec T_p)-\vec S$ 
of the star image can also be predicted. In the uv-plane, the predicted image of the star is at
\begin{equation}
	\vec{I_{p,uv}}=\vec{T_p}+u_p\vec{e_{\phi D}}+v_p\vec{e_{\theta D}}=\frac{1}{\vec{T_p}\cdot\vec{I_p}}\vec{I_p}\;\mathrm{.}
\end{equation}
The last equality results from $\vec{I_{p,uv}}=|\vec{I_{p,uv}}|\vec{I_p}=\vec{I_p}\cos\psi$ where $\psi:=\angle(\vec{T_p},\vec{I_p})$, see also Fig. \ref{geometry_fig}.\\
For optimal pointing parameters, the predicted star image coordinates $(u_p, v_p)$ in the uv-plane, i.e. 
	\begin{equation}
	\label{model_uv}
		\begin{array}{l}
			u_p=\frac{1}{\vec{T_p}\cdot\vec{I_p}}\vec{I_p}\cdot\vec{e_{\phi D}}\\
			v_p=\frac{1}{\vec{T_p}\cdot\vec{I_p}}\vec{I_p}\cdot\vec{e_{\theta D}}\;\mathrm{,}	
		\end{array}		
	\end{equation}
match the $(u,v)$ coordinates inferred from CCD images (Eq. \ref{measurement_uv}) within deviations expected in the likelihood model. 
\subsection{Likelihood model and treatment of measurement outliers}
The likelihood function for the parameters $\vec p$ and $\vec\pi$ given 
the values of $\vec{u}:=(u_1,\dots, u_N)$ and $\vec{v}:=(v_1,\dots, v_N)$ factorizes like
\begin{equation}
	\tilde{\mathcal{L}}(\vec{p},\vec{\pi}|\vec{u},\vec{v})=\prod_{i=1}^N\mathcal{L}(\vec{p},\vec{\pi}|u_i)\,\mathcal{L}(\vec{p},\vec{\pi}|v_i)\,\mathrm{.}
	\label{likelihood_model}
\end{equation}
For example, consider the case where a Gaussian likelihood model given measured $u_i$ and predicted $u_{ip}(\vec p)$,
	      \begin{equation}
		      \mathcal{L}(\vec{p},\vec{\pi}|u_i)=\frac{1}{\sqrt{2\pi\sigma^2}}\exp\left(\frac{(u_i- u_{ip}(\vec{p}))^2}{2\sigma^2}\right)\;\mathrm{,}
	      \end{equation}
and respectively for $\mathcal{L}(\vec p,\vec\pi|v_i)$ is assumed. In this case, the full set of nuisance parameters is $\vec{\pi}=(\sigma,\alpha)$ where $\sigma$ describes the measurement error of 
$u$ and $v$ and a maximum likelihood optimization can be interpreted geometrically as being 
equivalent to the minimization of the sum of squared Euclidean distances $\sum_{i=1}^N\left(u_i- u_{ip}\right)^2 + \left(v_i- v_{ip}\right)^2$. However, the Gaussian likelihood model 
is very sensitive to measurement outliers which bias the optimal parameters or lead to convergence problems. 
Measurement outliers can efficiently be described when the Gaussian likelihood function is replaced with a Student likelihood function. 
The number of degrees of freedom $\nu$ of the likelihood function is then a nuisance parameter, 
additionally to $\sigma$ and $\alpha$.
\subsection{Bayesian analysis}
CTbend performs a Bayesian analysis to find the optimal pointing and nuisance parameters as mean values of the respective posterior distributions. 
The Bayesian analysis allows for the natural inclusion of prior 
information on the parameters. For example, many pointing model parameters in $\vec p$ are known to be small in absolute value while, for example, the nuisance parameters $\nu$ and $\sigma$ must 
be positive. These parameter constraints can lead to difficult 
convergence problems when, e.g., a maximum likelihood optimization is performed. Therefore,
a Markov Chain Monte Carlo (MCMC), based on an implementation which is described in \cite{pymc3}, is performed. 
Table \ref{prior_table} lists a possible choice for prior distributions. Laplace priors are used for pointing model parameters. This choice is equivalent to a LASSO regularization of 
the model parameters \cite{bayeslasso}. A log-normal and exponential prior is used for the parameters $\sigma$ and $\nu$.
This ansatz is common in Bayesian analyses, see e.g. \cite{bda}, \cite{rethinking}. A fixed parameter is used for the CCD camera rotation angle $\alpha$. The priors listed in Tab. \ref{prior_table} 
are only an example, the functional form of the priors can be easily changed in CTbend.\\
Given the likelihood function (Eq. \ref{likelihood_model}) and prior distributions for the parameters, the posterior parameter distributions are derived by matching the measured and predicted 
values for $u$ and $v$ in a MCMC simulation. The analysis workflow is summarized in Fig. \ref{workflow_figure}. Eventually, the mean of the posterior bending parameter distributions is
used as respective point estimate.
\begin{table}
\begin{center}
\begin{tabular}{ |c|c c| c|}
	\hline
	Description & $\Delta\phi_{pk}$ & $\Delta\theta_{pk}$ & $k$\\
	\hline
	Zero offsets & IA & IE & 0\\
	Azimuth/elevation axis non-perpendicularity & $\mathrm{NPAE}\tan\theta$ & 0 & 1 \\
	Azimuth axis north-south misalignment & $-\mathrm{AN}\tan\theta\sin\phi$ & $\mathrm{AN}\cos\phi$ & 2 \\
	Azimuth axis east-west misalignment & $-\mathrm{AW}\tan\theta\cos\phi$ & $-\mathrm{AW}\sin\phi$ & 3\\
	Tube flexure & 0 & $\mathrm{TF}\cos\theta$ & 4\\
	Azimuth centering error (cos component)& $\mathrm{ACEC}\cos\phi$ & 0 & 5 \\
	Azimuth centering error (sin component)& $\mathrm{ACES}\sin\phi$ & 0 & 6 \\
 \hline

\end{tabular}
	\caption{Terms $(\Delta\phi_{pk},\,\Delta\theta_{pk})$, of a basic 8-parameter pointing model $(\Delta\phi_p=\sum_{k=0}^6\Delta\phi_{pk},\,\Delta\theta_p=\sum_{k=0}^6\Delta\theta_{pk})$. 
	The nomenclature for the parameters is adopted from \cite{tpoint} where also more details regarding the interpretation of the respective terms can be found. 
	The orientation refers to an azimuth coordinate system where south is at $\phi=0$ and east is at $\phi=-90^\circ$.}

\label{example_model_table}
\end{center}
\end{table}
\begin{table}
\begin{center}
\begin{tabular}{ |c|c|c|c|}
 \hline
	Description & Prior distribution & Mean & Standard deviation\\
	\hline
	Bending model parameters & Laplace & $0$ & $10^\circ$\\
	Measurement error ($\sigma$)& Log-normal & $\tan(\xi)$ & $\tan(5\xi)$\\
	Degrees of freedom ($\nu$) & Exponential & $10$ & - \\
	CCD rotation angle ($\alpha$) & Fixed/Dirac prior & $0$ \\
 \hline

\end{tabular}
	\caption{Example prior distributions and parameters. Bending model parameters refer to all parameters listed in Tab. \ref{example_model_table}. 
	The measurement error, the number of degrees of freedom and the CCD rotation angle are nuisance parameters of the likelihood function given by Eq. \ref{likelihood_model}.}
	\label{prior_table}
\end{center}
\end{table}
\section{Conclusion and outlook}
\label{outlook_section}
CTbend is a standalone implementation of a framework to model pointing corrections for Cherenkov telescopes. 
A flexible choice of parametrizations for bending models is possible. A simple TPoint model was discussed but extensions of this model or completely different models, such as full Fourier models, 
are easily possible. Further progress is expected when more real data from different Cherenkov telescopes is processed in the future. Questions of model selection, 
e.g. based on information criteria \cite{aic}, 
are to be discussed in this context based on real data. Further open questions regard e.g. the principal limitations of the precision of pointing models 
and the extension of pointing models to also include time-dependent effects, e.g. in a hierarchical model (see e.g. \cite{bda}) for seasonal effects. 
Also, the interplay between an advanced understanding of pointing models and the automatic condition monitoring of telescope structures in telescope arrays with a large number of 
individual telescopes can be subject to further studies.
\\

\textbf{Acknowledgments:} GS and US acknowledge the support by the German Ministry for Education and Research (BMBF).

%
%
%


\begin{thebibliography}{99}
	\bibitem{hess}Aharonian et al. (2006), Observations of the Crab nebula with HESS, A\&A, 457 (3)
	\bibitem{taiga}Tluczykont et al. (2014), The HiSCORE concept for gamma-ray and cosmic-ray astrophysics beyond 10 TeV, Astroparticle Physics, 56
	\bibitem{magic}Aleksic et al. (2012), Astroparticle Physics, 35 (7)
	\bibitem{veritas}Holder et al. (2006), The first VERITAS telescope, Astroparticle Physics, 25
	\bibitem{cta}CTA Consortium (2019), Science with the Cherenkov Telescope Array, World Scientific
	\bibitem{alma}Mangum et al. (2006), Evaluation of the ALMA Prototype Antennas, Publications of the Astronomical Society of the Pacific, 118 (847)
	\bibitem{meeks}Meeks et al. (1968), The pointing calibration of the Haystack antenna, IEE Transactions on Antennas and Propagation, 16
	\bibitem{bayeslasso}T. Park and G. Casella (2008), The Bayesian Lasso, Journal of the American Statistical Association, 103 (482)
	\bibitem{pymc3}Salvatier et al. (2016), Probabilistic programming in Python using PyMC3, PeerJ Computer Science, 2
	\bibitem{bda}Gelman et al. (2014), Bayesian Data Analysis, CRC Press, 3rd edition
	\bibitem{rethinking}R. McElreath (2016), Statistical Rethinking, CRC Press
	\bibitem{tpoint}P. T. Wallace (1994), TPoint - Telescope Pointing Analysis System, Starlink user note
	\bibitem{aic}H. Akaike (1974), A new look at the statistical model identification, in IEEE Transactions on Automatic Control, 19 (6)
\end{thebibliography}
\end{document}